\documentclass[seceq,preprint]{ptptex}

\usepackage{epsfig,multicol,bbm,bm}

\newcommand{\id}{{\mathbbm 1}}

\def\slash#1{\not\!#1}

\title{Characters of Lattice Fermions Based on the Hyperdiamond Lattice}

\title{
Characters of Lattice Fermions Based on the Hyperdiamond Lattice%
}

\author{
Taro \textsc{Kimura}$^{1}$%
and Tatsuhiro \textsc{Misumi}$^{2}$}

\inst{
$^{1}$Department of Basic Science, University of Tokyo, Tokyo 153-8902, Japan\\
$^{2}$Yukawa Institute for Theoretical Physics, Kyoto University, Kyoto 606-8502, Japan\\
}

\abst{
We study minimal-doubling fermion actions on hyperdiamond and deformed-hyperdiamond lattices, with emphasis on the real-space construction of them and Lorentz covariance of excitations from fermion poles.
We propose the improved spatial construction of Creutz fermion action on a deformed hyperdiamond lattice, and discuss conditions for a hyperdiamond-lattice action to produce Lorentz-covariant excitations from poles of fermion propagators. 
It is pointed out that the non-nearest-site hoppings are essential for the correct excitations.
We also propose a class of minimal-doubling actions defined on a deformed hypercubic lattice as a generalization of Creutz-type actions.
In addition we introduce a two-parameter class of Wilczek-type minimal-doubling actions.
}

\preprintnumber[3cm]{
YITP-09-41}

\begin{document}
\maketitle

\section{Introduction}
Recently, Creutz\cite{Creutz:2007af} and Bori\c{c}i\cite{Borici:2007kz} have proposed a two-parameter class of fermion actions called ``Creutz fermion'', inspired by the relativistic condensed matter system, graphene\cite{neto:109}.
This fermion is defined on the hyperdiamond lattice distorted by two parameters ($B, C$) and includes the non-nearest hopping terms.
What is notable about ``Creutz fermion'' is that it has desirable properties for the lattice simulation such as locality, chiral symmetry and the minimal number of fermion doubling.
Among them, the minimal fermion doubling is the most outstanding characteristic of this fermion.
As is well-known, although there exist only two (or three) light quarks in QCD, Nielsen-Ninomiya's no-go theorem\cite{Nielsen:1980rz,Nielsen:1981xu,Nielsen:1981hk} states that the lattice fermion with chiral symmetry and other common features inevitably yields degrees of freedom of multiple number of two in a continuum limit. 
On the other hand, the lattice fermions which bypass the no-go theorem such as domain-wall fermion\cite{Kaplan:1992bt,Furman:1994ky} and overlap fermion\cite{Ginsparg:1981bj, Neuberger:1998wv} demand an expensive numerical task.    
Therefore the chiral-symmetric fermion including only the minimal number of doubling, such as Creutz fermion, will be much faster and more useful in the simulation since the two fermion degrees of freedom can be directly interpreted as the two light quarks in lattice QCD simulation.

However it was pointed out \cite{Bedaque:2008xs} that Creutz action
lacks sufficient discrete symmetry to prohibit relevant and marginal
operators to be generated through the loop corrections which are serious
obstacles for a good continuum limit in the lattice simulation
\cite{Capitani:2009yn,Capitani:2010nn}. 
In particular, such a problem for the lattice action, which is
equivalent to Bori\c{c}i action, has already investigated\cite{Alonso:1987bb}.
In Ref.~\citen{Bedaque:2008jm} the authors show that if the non-nearest hopping terms are dropped with the parameters chosen to $B=1/\sqrt{5}$, $C=1$ in Creutz action, the requisite discrete symmetry of cyclic group $\mathbb{Z}_{5}$ recovers, although the modified action yields an unphysical excitation from the pole of propagator (or mutilated pole \cite{Celmaster:1982ht,Celmaster:1983jq,Drouffe:1983kq}).
They also construct a simple fermion action on a hyperdiamond lattice including only the nearest-neighbor hoppings.  
However, it is argued in \citen{Bedaque:2008jm} that although this fermion action has the sufficient discrete symmetry of alternating group $\mathfrak{A}_{5}\supset \mathbb{Z}_{5}$, it yields more than minimal number of doublers.

In this paper we investigate fermion actions on hyperdiamond and deformed-hyperdiamond lattices, with emphasis on the real-space construction of them and Lorentz-covariant excitations from poles of propagators, then obtain a generalized class of Creutz-type minimal-doubling actions on a deformed hypercubic lattice.
Firstly we propose the spatial construction of Creutz fermion action on a deformed hyperdiamond lattice,
which is an improved version of that in \citen{Bedaque:2008jm}.
Secondly we study conditions for a hyperdiamond-lattice action to produce Lorentz-covariant excitations from fermion poles. 
It is pointed out that the non-nearest neighbor hoppings in terms of sites are essential for the correct excitations.
Then we propose a class of minimal-doubling fermion actions defined on a deformed hypercubic (rhombus) lattice as a generalization of Creutz-type actions, where the link variables are easily introduced.
We also introduce a two-parameter class of Wilczek-type minimal-doubling
actions, which will be the simplest form of Creutz-type lattice action.

In Sec.~\ref{sec:Creutz} we briefly review the hyperdiamond lattice and
Creutz fermion.
In Sec.~\ref{sec:hyperdiamond} we discuss the spatial construction of Creutz action and investigate the conditions for the correct fermionic excitations. 
In Sec.~\ref{sec:examples} we study examples of hyperdiamond-lattice fermions and propose a related action ``Appended Creutz action''.
In Sec.~\ref{sec:rhombus} we generalize Creutz-type and Wilczek-type actions to classes of actions on deformed hypercubic lattices.
Section \ref{sec:summary} is devoted to a summary and discussion.

\section{Creutz fermion}\label{sec:Creutz}
We now consider the minimal-doubling action proposed by
Creutz, which is called Creutz
action\cite{Creutz:2007af,Borici:2007kz} and also Bori\c{c}i-Creutz
action in \citen{Bedaque:2008jm}.
The action is related to the hyperdiamond lattice, which is the higher
dimensional generalization of graphene system.
General dimensional aspects of the hyperdiamond
lattice and lattice fermions defined on them are discussed in \citen{KM:2009lf}.

The four-dimensional hyperdiamond lattice is constructed with five bond
vectors,
\begin{equation}
\begin{array}{c}
  \bm{e}^1=\frac{1}{4}(\sqrt{5},\sqrt{5},\sqrt{5},1), \quad
   \bm{e}^2=\frac{1}{4}(\sqrt{5},-\sqrt{5},-\sqrt{5},1), \\
  \bm{e}^3=\frac{1}{4}(-\sqrt{5},-\sqrt{5},\sqrt{5},1), \quad
   \bm{e}^4=\frac{1}{4}(-\sqrt{5},\sqrt{5},-\sqrt{5},1), \\
  \bm{e}^5=(0,0,0,-1), \label{hyperdiamond_bond}
\end{array}
\end{equation}
satisfying
\begin{equation}
 \bm{e}^\mu \cdot \bm{e}^\nu = \left\{ \begin{array}{ccc}
				   1  & \mbox{for} & \mu =
				    \nu \\
				   \cos \theta = -1/4  & \mbox{for} & \mu \not= \nu \\
					    \end{array} \right..
\end{equation}
The spatial translation symmetry of the hyperdiamond lattice is characterized by
primitive vectors $\bm{d}_\mu\,\,(\mu=1,2,3,4)$
defined as
\begin{equation}
 \bm{d}_\mu = \bm{e}^\mu - \bm{e}^5 \quad \mbox{for} \quad \mu = 1,
  \cdots, 4. \label{primitive01}
\end{equation}
Then an angle $\eta$ between them is given by
\begin{equation}
 \cos \eta =
  \frac{\bm{d}_\mu\cdot\bm{d}_\nu}{\left|\bm{d}_\mu\right|\left|\bm{d}_\nu\right|}
  = \frac{1}{2}. \label{diamond_angular}
\end{equation}
This is a common property for any dimensional hyperdiamond lattices\cite{KM:2009lf}.

An important property of the hyperdiamond lattice is its sublattice
structure, such that it consists of two kinds of sites.
These sublattices are called $L$-node and $R$-node\footnote{In the case
of the graphene system, these sublattice structure is often represented by
$A$ and $B$ sites.}, whose positions
are labeled by $x_L = \sum_{\mu=1}^4 x_\mu \bm{d}_\mu$ and  $x_R
= \sum_{\mu=1}^4 x_\mu \bm{d}_\mu + \bm{e}^5$, but this node-index is
often omitted in the following discussion because only positions of unit
cells are important for a lattice action.
This sublattice structure corresponds to chirality of fermions, and is in common with the honeycomb lattice $(d=2)$ and the diamond lattice $(d=3)$.
That means, if poles of the Dirac operator are arranged to construct the hyperdiamond lattice in momentum space, the number of doublers is only two.
This is a strategy to obtain the hyperdiamond-type minimal-doubling fermion, and actually done in Creutz's original
paper\cite{Creutz:2007af}.

Then let us show how four dimensional generalization
of the graphene system is considered.
To investigate the graphene system, we often use the tight-binding model
on a honeycomb lattice for $\pi$-electrons of carbon atoms.
The Hamiltonian in momentum space is given by
\begin{equation}
 H(p) = K \left(\begin{array}{cc}
	   0 & z(p) \\ z^*(p) & 0
		\end{array}\right)
\end{equation}
where $K$ is a hopping amplitude and an off-diagonal component is defined as
$z(p)=1+e^{ip_1}+e^{ip_2}$.
This Hamiltonian is associated with a conventional anti-hermitian Dirac
operator in lattice field theory as $H=\sigma_3 D$, and we now use
non-orthogonal coordinates defined by primitive vectors, $p_\mu =
\bm{d}_\mu\cdot p$.

To consider four-component Dirac spinor in four dimensions, 
a complex number $z(p) \in \mathbb{C}$ is generalized to a quaternion
$z = c_0+i\vec{c}\cdot\vec\sigma \in \mathbb{H}$. 
In the original paper\cite{Creutz:2007af}, 
the associated Dirac operator is given by
\begin{eqnarray}
  D(p)\  = 
 && (\sin p_1 + \sin p_2 - \sin p_3 - \sin p_4) i \gamma_1 \nonumber \\
 &+& (\sin p_1 - \sin p_2 - \sin p_3 + \sin p_4) i \gamma_2 \nonumber \\
 &+& (\sin p_1 - \sin p_2 + \sin p_3 - \sin p_4) i \gamma_3 \nonumber \\
 &+& B (4C - \cos p_1 - \cos p_2 - \cos p_3 - \cos p_4) i \gamma_4.
 \label{Creutz_Dirac}
\end{eqnarray}
Gamma matrices are defined as $\gamma_i = \tau_1 \otimes
\sigma_i$, $\gamma_4 = \tau_2 \otimes \id$ and $\gamma_5 = \tau_3
\otimes \id$
where $\tau$'s and $\sigma$'s are both Pauli matrices acting on the
sublattice and internal spinor structure, respectively.
Here, as the graphene system, we use
non-orthogonal coordinates, $p_\mu = \bm{a}_\mu\cdot p$, where
$\bm{a}_\mu$ are primitive vectors of the lattice for Creutz action.
However in Sec.~\ref{sec:hyperdiamond} we will show they are different from those defined in (\ref{primitive01}).

This Dirac operator possesses two poles at $p=\pm(\tilde p, \tilde p, \tilde
p, \tilde p)$ with $\cos \tilde p =C$, if and only if the lattice
parameter $C$ satisfying $1/2 < C < 1$, to suppress extra poles such as
$(\tilde p, \tilde p, \tilde p, \pi - \tilde p)$.
Although these two poles induce physical Dirac fermions, in the case
$C=1$ they are reduced to only one cut rather than a pole, and thus it turns
out to be unphysical.
To show this, we expand the Dirac operator (\ref{Creutz_Dirac}) around the pole as $p_\mu = \tilde p + q_\mu$,
\begin{eqnarray}
  D(p)\ = && C (q_1 + q_2 - q_3 - q_4) i \gamma_1 \nonumber \\
 &+& C (q_1 - q_2 - q_3 + q_4) i \gamma_2 \nonumber \\
 &+& C (q_1 - q_2 + q_3 - q_4) i \gamma_3 \nonumber \\
 &+& BS (q_1 + q_2 + q_3 + q_4) i \gamma_4 + {\cal O}(q^2)
 \label{Creutz_Dirac02}
\end{eqnarray}
with $S = \sin \tilde p$.
This Dirac operator behaves as $i \vec{\gamma} \cdot \vec{k}$
around $p_\mu = 0$ with $S=0\ (C=1)$, and thus it has been shown that
covariance of this fermion is broken.
This unphysical fermion is known as a mutilated fermion, often found in some attempts on nonhypercubic lattices\cite{Celmaster:1982ht,Celmaster:1983jq,Drouffe:1983kq}.

Since gamma matrices satisfy anti-commutation relations $\{\gamma_\mu,
\gamma_\nu\}=2\delta_{\mu\nu}$, coefficients of gamma matrices are
interpreted as those of a momentum represented by Euclidean coordinates.
Thus reciprocal vectors $\{\bm{b}^\mu\}$ giving momentum space basis are obtained by (\ref{Creutz_Dirac02}),
\begin{equation}
\begin{array}{c}
  \bm{b}^1=(C,C,C,BS), \quad
   \bm{b}^2=(C,-C,-C,BS), \\
  \bm{b}^3=(-C,-C,C,BS), \quad
   \bm{b}^4=(-C,C,-C,BS).
\end{array} \label{reciprocal01}
\end{equation}
These vectors characterize the translation structure in momentum space,
thus they can be interpreted as ``primitive vectors'' of momentum space. 

To consider a situation such that poles construct the exact hyperdiamond
lattice, as primitive vectors (\ref{diamond_angular}), it is imposed
in Ref.\ \citen{Creutz:2007af} that an angle $\xi$ between reciprocal vectors
satisfies $\cos \xi = 1/2$.
Here it is given by
\begin{equation}
 \cos \xi =
  \frac{\bm{b}^\mu\cdot\bm{b}^\nu}{|\bm{b}^\mu||\bm{b}^\nu|}
 = \frac{B^2S^2 - C^2}{B^2S^2 + 3 C^2}.
 \label{reciprocal_angle}
\end{equation}
With adjusting lengths of reciprocal vectors, Creutz chose two
parameters, $C=\cos (\pi/5)$ and $B=\sqrt{5}\cot (\pi/5)$.
On the other hand, the orthogonal condition
$\bm{b}^\mu\cdot\bm{b}^\nu=0$ gives $BS=C$ applied in
Ref. \citen{Borici:2007kz}.

We now remark an important property of reciprocal vectors to consider
the lattice structure in real space.
Because an arbitrary momentum is represented by $p = \sum_{\mu=1}^4
\left(p\cdot\bm{a}_\mu\right) \bm{b}^\mu$, associated primitive vectors
are determined by a relation $\bm{a}_\mu\cdot\bm{b}^\nu=\delta_\mu^{\nu}$.
We will discuss real-spatial construction of
Creutz fermion with this relation in Sec.~\ref{sec:hyperdiamond}.

\section{Hyperdiamond lattice fermion}\label{sec:hyperdiamond}

As discussed in the previous section, Creutz fermion was directly
constructed in momentum space.
Then we should consider its lattice construction in real space to
reproduce Creutz's Dirac operator (\ref{Creutz_Dirac}). 
The translation structure in momentum space is given by the
expanded Dirac operator (\ref{Creutz_Dirac02}), and provides the associated
real-spatial lattice structure.

In this section we introduce a real-spatial construction of Creutz fermion
on the deformed hyperdiamond lattice.
Then we show that the bond vectors of the lattice and the hopping vectors of the fermion fields should be determined by the reciprocal vectors Eq.~(\ref{reciprocal01}) consistently.
In this sense our construction is more reasonable than that in \citen{Bedaque:2008jm} although they seem similar to each other.
Based on this construction, we will give a parameter condition for
Creutz fermion to be defined on the exact hyperdiamond lattice in real space.
We will also discuss conditions for hyperdiamond-lattice fermions
to yield only physical or Lorentz-covariant excitations.

\subsection{Creutz fermion on hyperdiamond lattice}\label{sec:hyperdiamond_Creutz}

To obtain a four-component Dirac fermion on the four dimensional
hyperdiamond lattice, we consider
two-component chiral spinors, left-handed $\phi$ and right-handed
$\bar{\phi}$ located on $L$-node, and right-handed $\chi$ and
left-handed $\bar{\chi}$ located on $R$-node. 
Here we note that $\phi$ and $\bar{\phi}$ are not hermite conjugate, but
independent degrees of freedom. 
This configuration of the lattice fermions will play an important role
on the discussion in section \ref{sec:features}.

Here we define ``spinor vectors'', which appear as the coefficient vectors of the gamma matrices in the action.
The spinor vectors with parameters $B$ and $C$ are given by
\begin{equation}
\begin{array}{c}
  \bm{s}^1 = (1,1,1,B) , \quad  \bm{s}^2 = (1,-1,-1,B) , \\
  \bm{s}^3 = (-1,-1,1,B) , \quad  \bm{s}^4 = (-1,1,-1,B) , \\
  \bm{s}^5 = (0,0,0,-4BC). \label{twisted_bond}
\end{array}
\end{equation}
The action is given by
\begin{eqnarray}
  S_{\mathrm{C}}
 & = & \frac{1}{2} \sum_x \Big[ \sum_{\mu=1}^4 \left(
 \bar{\phi}_{x-\bm{a}_\mu} \Sigma \cdot \bm{s}^\mu \chi_x 
 - \bar{\chi}_{x+\bm{a}_\mu} \Sigma \cdot \bm{s}^\mu \phi_x
 \right) \nonumber \\
 && + \bar{\phi}_x \Sigma \cdot \bm{s}^5 \chi_x 
 - \bar{\chi}_x \Sigma \cdot \bm{s}^5 \phi_x \nonumber \\
 && + \sum_{\mu=1}^4 \left(
 \bar{\chi}_{x-\bm{a}_\mu} \bar \Sigma \cdot \bm{s}^\mu \phi_x 
 - \bar{\phi}_{x+\bm{a}_\mu} \bar \Sigma \cdot \bm{s}^\mu \chi_x
 \right) \nonumber \\
 && + \bar{\chi}_x \bar\Sigma \cdot \bm{s}^5 \phi_x 
 - \bar{\phi}_x \bar\Sigma \cdot \bm{s}^5 \chi_x \Big]. \label{Creutz_action}
\end{eqnarray}
where the hopping vectors $\bm{a}_{\mu}$ (or primitive vectors) will be
determined to be consistent with the reciprocal vectors in the momentum
space afterward.
Here the fourth components of spinor matrices are twisted as
$\Sigma=(\vec{\sigma},-1)$, $\bar{\Sigma}=(\vec{\sigma},1)$.
It is notable that hopping terms of $\phi_x \to \bar\chi_{x-\bm{a}_\mu}$ and
$\chi_x \to \bar\phi_{x+\bm{a}_\mu}$ in (\ref{Creutz_action}) represent
non-nearest site interactions.

Thus we obtain Creutz's Dirac operator (\ref{Creutz_Dirac}) by
considering Fourier transformation of the lattice action
(\ref{Creutz_action}) as
\begin{equation}
 S_{\mathrm{C}} = \int \frac{d^4 p}{(2\pi)^4}\ \bar{\psi}_p D(p) \psi_p
\end{equation}
with a Dirac spinor
\begin{equation}
 \psi_p = \left( \begin{array}{c} \phi_p \\ \chi_p \end{array} \right), \quad
 \bar{\psi}_p = \left( \bar{\phi}_p \quad \bar{\chi}_p \right).
\end{equation}

In this construction the reciprocal vectors are obtained as (\ref{reciprocal01}).
Since primitive and reciprocal vectors satisfy
$\bm{a}_\mu\cdot\bm{b}^\nu=\delta_\mu^{\nu}$,
the associated primitive vectors, or the hopping vectors of the fermion fields, for Creutz fermion are given by
\begin{equation}
\begin{array}{c}
  \bm{a}_1=\frac{1}{4C}(1,1,1,\frac{C}{BS}), \quad
   \bm{a}_2=\frac{1}{4C}(1,-1,-1,\frac{C}{BS}), \\
  \bm{a}_3=\frac{1}{4C}(-1,-1,1,\frac{C}{BS}), \quad
   \bm{a}_4=\frac{1}{4C}(-1,1,-1,\frac{C}{BS}).
\end{array} \label{primitive02}
\end{equation}
Since primitive vectors of the (distorted) hyperdiamond lattice are
given by (\ref{primitive01}), bond vectors $\{\bm{e}^\mu\}$ are
obtained by introducing another free parameter $f$ ($f>0$),
\begin{eqnarray}
 \bm{e}^\mu &=& \bm{a}_\mu + \bm{e}^5 \quad \mbox{for} \quad \mu=1,2,3,4,
  \nonumber \\
 \bm{e}^5 &=& (0,0,0,-f).
 \label{twisted_bond02}
\end{eqnarray}
These vectors are identified with the bond vectors of the hyperdiamond
lattice on which the action is defined, while the spinor vectors
$\{\bm{s}^\mu\}$ in the action (\ref{Creutz_action}) are just related to
the lattice structure indirectly.

\begin{figure}[tbp]
 		\begin{minipage}{0.5\hsize}
 			\begin{center}
 				\includegraphics[height=2.7cm]{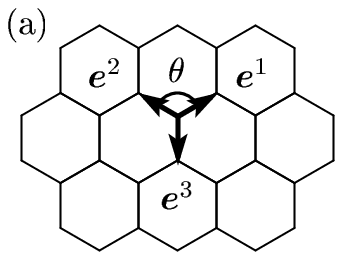}
 			\end{center}
 		\end{minipage}%
 		\begin{minipage}{0.5\hsize}
 			\begin{center}
 				\includegraphics[height=2.7cm]{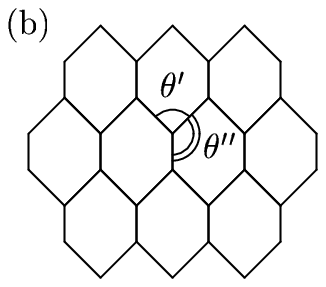}
 			\end{center}
 		\end{minipage}
  \caption{Two dimensional analogue of the lattice deformation: (a) the two
 dimensional regular diamond (honeycomb) lattice and (b) the distorted
 lattice with $\bm{e}^{3}$ direction specified $(\theta''>\theta>\theta')$.}
  \label{honeycomb_lattice}
\end{figure}

For general values of the parameters, the hyperdiamond lattice is deformed such that it is elongated in $\bm{e}^5$ direction as shown in Fig.~\ref{honeycomb_lattice}.
As the case of the exact hyperdiamond lattice (\ref{diamond_angular}),
an angle between the primitive vectors (\ref{primitive02}) is given by
\begin{equation}
 \cos \eta = \frac{C^2 - B^2 S^2}{C^2 + 3B^2S^2}.
\end{equation}
We note that it is related to (\ref{reciprocal_angle}) by exchanging $C
\leftrightarrow BS$. 
Then if we choose deformation parameters as $f = 1/(\sqrt{5}C)$ and
$BS=C/\sqrt{5}$, the angles become $\cos \eta = 1/2$ and $\cos \theta =
\bm{e}^\mu \cdot \bm{e}^\nu/(|\bm{e}^\mu||\bm{e}^\nu|)= -1/4$, and thus
Creutz fermion is defined on the exact hyperdiamond lattice. 
We will call this condition ``hyperdiamond condition''.
In this case, the angle between reciprocal vectors becomes $\cos
\xi = -1/4$.
Furthermore, the Creutz and Bori\c{c}i's conditions such that in
momentum space poles are located on the exact hyperdiamond lattice and
on the orthogonal lattice imply the associated angles become
\begin{equation}
 \cos \eta = \left\{ \begin{array}{cl}
	      -1/4 & \mbox{(Creutz)} \\ 0 & \mbox{(Bori\c{c}i)}
		     \end{array}\right. .
\end{equation}
This means, in the sense of real and momentum spaces,  the hyperdiamond
condition and the Creutz condition are dual, and the Bori\c{c}i
condition is self-dual.

Here we remark discrete symmetry of Creutz fermion.
With general parameters, this distorted lattice and also the action on this have only $\mathfrak{S}_{4} \not\supset \mathbb{Z}_{5}$ symmetry which is not the sufficient discrete symmetry for a good continuum limit \cite{Bedaque:2008jm}.
In the case of the hyperdiamond condition, this distorted lattice becomes the regular $\mathfrak{S}_{5} \supset \mathbb{Z}_{5}$ symmetric hyperdiamond lattice.
However, even if the hyperdiamond lattice becomes exact, the non-nearest hopping terms reduce the discrete symmetry of the action to $\mathfrak{S}_{4}$. 
Thus the physical Creutz fermion cannot have the requisite discrete symmetry to prohibit the redundant operators. 
It indicates a general property that the minimal-doubling lattice fermion lacks the sufficient discrete symmetry for a continuum limit on hyperdiamond lattices.

\subsection{Conditions for physical hyperdiamond lattice fermions}\label{sec:features}

We have seen that physical fermionic excitations are obtained from minimal-doubling poles of Creutz's Dirac operator, although the action lacks sufficient discrete symmetry.
Here we investigate conditions for a hyperdiamond-lattice action to produce Lorentz-covariant excitations from poles of fermion propagators, which actually Creutz fermion satisfies.
We then argue ``minimal-doubling'' on the hyperdiamond lattice is incompatible with the sufficient discrete symmetry of the action for a good continuum limit since the above conditions, which we will call ``physicality conditions'', cannot be satisfied without lowering the symmetry of the action.

{\it Non-nearest neighbor hoppings.}
We have shown that Creutz action can be regarded as being defined on the hyperdiamond lattice.
From the viewpoint of this real-spatial interpretation, the action (\ref{Creutz_action}) contains the non-nearest neighbor hoppings which sometimes lead to breaking of locality.
However in this case, non-nearest neighbor hoppings are actually
{\it non-nearest} interactions in the sense of sites, but {\it nearest} for unit cells. 
The hoppings based on the vectors $\bm{a}_{\mu}$ in Eq.~(\ref{primitive02}) stand for nearest neighbor hoppings between the unit cells, not the sites.
Thus the locality is not broken in the continuum limit of
Creutz action as seen in \citen{Cichy:2008gk,Cichy:2008nt}.
Here we denote the nearest (non-nearest) hoppings in the sense of sites as ``nearest-site (non-nearest-site) hoppings''.
As follows, we will show the non-nearest-site hoppings are required for Lorentz-covariant or physical excitations of fermions on hyperdiamond lattices although the nearest-site hoppings are enough for a correct form of a Euclidean Dirac operator, namely anti-hermitian Dirac operator.


\begin{figure}[tbp]
			\begin{center}
				\includegraphics[height=2.7cm]{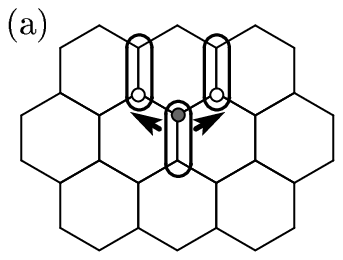} \quad
				\includegraphics[height=2.7cm]{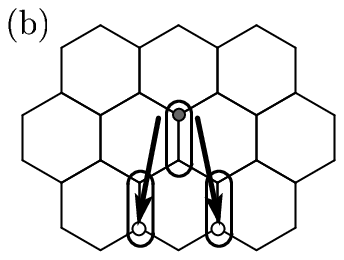} \quad
				\includegraphics[height=2.7cm]{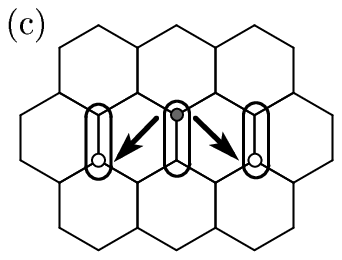}
			\end{center}
 \caption{$L \to R$ hoppings to nearest unit cells\cite{KM:2009lf}: (a) forward (nearest
 neighbor site) hoppings, (b) backward (non-nearest neighbor site) hoppings, and (c) the remaining nearest unit cell hoppings which are not included by Creutz action. $L$-node and $R$-node are denoted by shaded and open circles, respectively. Unit cells are encircled.}
 \label{honeycomb_nonnearest}
\end{figure}

For nearest-site $L \to R$ hoppings on the hyperdiamond lattice, a forward hopping as $x \to
x+\bm{a}_\mu$ is allowed but a backward hopping $x \to x - \bm{a}_\mu$
is not as shown in Fig.~\ref{honeycomb_nonnearest}, and they are
inverted in the case of $R \to L$ hoppings.
Now let us consider a hyperdiamond-lattice action only with nearest-site hoppings as discussed in \citen{Bedaque:2008jm}.
Although either of $L \to R$ or $R \to L$ hopping corresponds to a
non-hermitian operator as $i(\partial_\mu - 1)$ or $i(1 - \partial_\mu^\dagger)$, we can make the Euclidean Dirac operator anti-hermitian (namely a correct form) by including both of $L \to R$ and $R \to L$ nearest-site hopping terms.
However in this case, only either of $e^{ip_\mu}$ or $e^{-ip_\mu}$ appears in the coefficients of spinor matrices such as $\Sigma$ and $\bar{\Sigma}$ in the momentum space.
Since these coefficients are complex numbers, we should introduce two
anti-hermitian basis, $i\gamma$ and $\gamma\gamma_5$, in order to expand this kind of the operator by gamma matrices. (Here the chirality of the Dirac operator is not broken since both of $i\gamma$
and $\gamma\gamma_5$ anticommute with $\gamma_5$.)
Then this kind of the Dirac operator is given by
\begin{equation}
D(p)\,=\, \sum_{\mu}i\gamma_{\mu}F_{\mu}(p)+\sum_{\mu}\gamma_{\mu}\gamma_{5}G_{\mu}(p)
\label{generalD}
\end{equation}
where $F_{\mu}(p)$ and $G_{\mu}(p)$ are independent {\it real} functions in general.
This type of Dirac operator is actually proposed in
\citen{Bedaque:2008xs}, as the generally chiral symmetric operator.
In this operator $i\gamma$-terms and $\gamma\gamma_5$-terms are regarded
as ``vector'' and ``axial-vector'' functions, respectively, and thus
Nielsen-Ninomiya's no-go theorem cannot be applied to this kind of Dirac
operator because it is based on Poincar\'e-Hopf theorem for ``either''
of vector or axial-vector functions, not for both of them.
Therefore, there is no guarantee that the Dirac operator yields physical
poles of fermion propagator and the number of poles becomes even.
Thus it indicates that the index of the Dirac operator including both of
$i\gamma$ and $\gamma\gamma_5$-terms is ill-defined.

Actually the operator including both of $i\gamma$-terms and $\gamma\gamma_5$-terms yields unphysical fermion doublers in general. (Even on hypercubic lattices, the difference operator including only either of forward or backward hoppings induces unphysical poles \cite{Aoki:2004}.)
On the other hand, as seen in Creutz action, we can expand the Dirac operator by only $i\gamma$-terms if we introduce hoppings to non-nearest neighbor sites but nearest unit cells.
Thus non-nearest-site interactions are required for constructing a physical Dirac operator, which produces only physical degrees of freedom on the hyperdiamond lattice.

Here let us note that hermiticity of the hyperdiamond-lattice Dirac operator without non-nearest hoppings, which is one of the necessary conditions for Nielsen-Ninomiya's theorem, does not imply the operator can be expanded by only either of ``vector'' or ``axial-vector'' functions, as is different from the hypercubic case.
Thus we may be able to claim that it is, in a sense, a counter example of Nielsen-Ninomiya's theorem since the associated Dirac operator includes both of $i\gamma$ and $\gamma\gamma_5$ terms although all the conditions for the theorem are satisfied.

As shown above, non-nearest neighbor interactions are
necessary for a physical fermionic mode on hyperdiamond lattices.
However this kind of the non-nearest hopping terms lower the discrete symmetry of the action \cite{Bedaque:2008jm}.
Therefore it seems impossible to construct a physical fermion action with the sufficient discrete symmetry on the hyperdiamond lattice as far as it is based on the initial setting of the field
configuration as discussed in section
\ref{sec:Creutz}, namely, two kinds of chiral fermions on
$L$-nodes and $R$-nodes, respectively.
It also means that the requisite discrete symmetry for a good continuum limit is incommensurate with physical minimal-doubling actions on hyperdiamond lattices in this configuration of fermion fields.  
This no-go property is first conjectured in \citen{Bedaque:2008jm}, and we discuss it systematically as shown above.

{\it Twisting spinor structure}.
To obtain a physical mode, Dirac operator must be expanded by either of $i\gamma$ or $\gamma\gamma_5$-terms.
The most important condition is including non-nearest-site hoppings, but we
now show some subsidiary conditions are required for constructing
a physical hyperdiamond lattice fermion.
To construct a physical Dirac operator, we should multiply some spinor matrices
depending on hopping directions.
The naive choice is $\sigma =(\vec{\sigma}, i)$ as proposed in
\citen{Bedaque:2008jm}.
On the other hand, in the case of Creutz action the fourth component of
spinor structure is twisted as $\Sigma = (\vec{\sigma}, -1)$, and the same
coefficients are applied to non-nearest-site hoppings.
Due to this twist, $\sin p$ is converted to $\cos p$ in the coefficient
of the corresponding gamma matrix, and thus Creutz's Dirac operator becomes anti-hermitian by twisting the spinor component.

Although the spinor structure of Creutz action is apparently unnatural, it
should be determined in order to construct an anti-hermitian Dirac operator.
It is expected
that a nontrivial spinor structure is related to an action based on a
non-Bravais lattice which possesses sublattice structure, e.g. staggered
fermion\cite{Susskind:1976jm}.

{\it Lattice deformation}.
Besides the above prescriptions to obtain physical modes, we need deformed 
``spinor vectors'' by elongating in one specific direction.
As discussed in Sec.~\ref{sec:hyperdiamond_Creutz}, these spinor vectors
cannot be interpreted as bond vectors of the lattice.
But we have shown they are related as follows: spinor vectors
$\{\bm{s}^\mu\}$ give the Dirac operator and also reciprocal vectors
$\{\bm{b}^\mu\}$ characterizing the translation symmetry in momentum space, then
primitive vectors $\{\bm{a}_\mu\}$ are derived from the condition
$\bm{a}_\mu\cdot\bm{b}^\nu=\delta_\mu^{\nu}$.
Thus two-parameter deformation of spinor vectors leads to the distorted
hyperdiamond lattice.
Although we have also shown that Creutz fermion can be constructed on the
exact hyperdiamond lattice, the action yields a cut on $(p_{1}, p_{2},
p_{3})=(0,0,0)$, not a pole when we apply the regular hyperdiamond
lattice bond vectors with the condition $C=1$ to spinor vectors. 
In this sense, the lattice deformation is also necessary for
physical poles although the discrete symmetry of the hyperdiamond lattice is broken.

\section{Examples}\label{sec:examples}

According to the three conditions discussed in section \ref{sec:features}, (i) Non-nearest-site hopping, (ii) Spinor twist, and (iii) Distortion of the lattice, we can consider eight kinds of fermions.
In this section we try to complete the classification of lattice fermions based on the hyperdiamond lattice in the aspects of the three features.

\subsection{BBTW fermion and Dropped Creutz fermion}\label{sec:BBTW}
Firstly let us consider hyperdiamond-lattice actions with the sufficient discrete symmetry for a good continuum limit. 
One of these fermion actions is proposed in \citen{Bedaque:2008jm}, which we will call BBTW action in this paper.
It is constructed with exact hyperdiamond spinor vectors and includes
only nearest-site hoppings with untwisted spinor structure vectors
$\sigma=(\vec{\sigma}, i)$ and $\bar{\sigma}=(\vec{\sigma}, -i)$ as
\begin{equation}
 S_{\mathrm{BBTW}} = \sum_x \Big[ \sum_{\mu=1}^4 \left( \bar{\phi}_{x-\bm{a}_\mu} \sigma \cdot \bm{s}^\mu \chi_x - \bar{\chi}_{x+\bm{a}_\mu} \bar{\sigma} \cdot \bm{s}^\mu \phi_x \right) + \bar{\phi}_x \sigma \cdot \bm{s}^5 \chi_x - \bar{\chi}_x \bar{\sigma} \cdot \bm{s}^5 \phi_x \Big],\label{BBTW_action}
\end{equation}
where spinor vectors $\bm{s}^\mu$ are those in (\ref{twisted_bond}) with
$B=1/\sqrt{5}$, $C=1$.
The hopping vectors $\bm{a}_{\mu}$ should be
obtained from the reciprocal vectors in the momentum space as discussed
in Sec.\ref{sec:hyperdiamond_Creutz}.
Taking Fourier transformation of the action, we obtain the associated Dirac operator represented as
\begin{equation}
 D(p) = i \sum_\mu^4 \left( \bm{s}^\mu \cdot \gamma \right) \sin p_\mu - \left( \sum_{\mu=1}^4 \bm{s}^\mu \cos p_\mu + \bm{s}^5 \right) \cdot \gamma \gamma_5 \label{BBTW_Dirac}
\end{equation}
The Dirac operator of BBTW action has at least seven spectral zeros at $p_\mu = 0$ and $p_1 = - p_2 = - p_3 = p_4 = \cos^{-1}(-2/3)$, etc.
Denoting the poles as $p_\mu = \hat{p}_\mu$, the operator is expanded
around the pole with momentum $p_\mu = \hat{p}_\mu + q_\mu$,
\begin{equation}
 D(p) = \sum_{\mu=1}^4 \Big[ i \left( \bm{s}^\mu \cdot \gamma \right) q_\mu \cos \hat{p}_\mu - \left( \bm{s}^\mu \cdot \gamma \gamma_5 \right) q_\mu \sin \hat{p}_\mu \Big] + {\cal O}(q^2). \label{BBTW_Dirac_exp}
\end{equation}
For the pole at $\hat{p}_\mu = 0$, the Dirac operator (\ref{BBTW_Dirac_exp}) becomes 
\begin{equation}
 D(p) = i \left( \bm{s}^\mu \cdot \gamma \right) q_\mu + {\cal O}(q^2).
\end{equation}
Thus this lattice action is reduced to the covariant Dirac form $i \left( \bm{s}^\mu \cdot \gamma \right) p_\mu \equiv i \slash{k}$, in the low energy region where we identify $k_\mu$ as the Cartesian momentum.
In the cases of other poles $\hat{p}_\mu \not=0$, however, the Dirac operator includes not only $i\gamma$-terms but $\gamma \gamma_5$-terms as seen in Eq.(\ref{BBTW_Dirac_exp}).
As a result, one cannot obtain a covariant form of excitation from these poles, but unphysical fermions.

Another lattice action with the sufficient discrete symmetry is obtained by modifying Creutz fermion.
To recover the sufficient discrete symmetry $\mathbb{Z}_5$ of Creutz
action to prohibit redundant operators, it was suggested in
\citen{Bedaque:2008jm} that one drop non-nearest hopping terms
and choose $B=1/\sqrt{5}$, $C=1$. 
Then we obtain another type of Creutz action,
\begin{equation}
 S_{\mathrm{dC}} = \sum_x \Big[ \sum_{\mu=1}^4 \left( \bar{\phi}_{x-\bm{a}_\mu} \Sigma \cdot \bm{s}^\mu \chi_x - \bar{\chi}_{x+\bm{a}_\mu} {\Sigma} \cdot \bm{s}^\mu \phi_x \right) + \bar{\phi}_x \Sigma \cdot \bm{s}^5 \chi_x - \bar{\chi}_x {\Sigma} \cdot \bm{s}^5 \phi_x \Big], \label{dCreutz_action}
\end{equation}
and then we call this action Dropped Creutz action.

In this case, the Dirac operator is almost the same as
(\ref{BBTW_Dirac}) but the definition of $i \gamma_4$ is modified as
$\gamma_5 \gamma_4$, which is also anti-hermitian.
As the case of BBTW fermion, both $\gamma$-terms and $\gamma\gamma_5$-terms are included, and thus Nielsen-Ninomiya's theorem cannot be applied to this lattice action.
Indeed the lattice fermion around $p_\mu=0$ is not written as a covariant form, $i \vec{\gamma} \cdot
\vec{k} + \gamma_5 \gamma_4 k_4$.


In BBTW action (\ref{BBTW_action}) any of the three conditions are not satisfied,
while two of them (i) Non-nearest-site hopping and (iii) Lattice deformation are not satisfied in Dropped Creutz action (\ref{dCreutz_action}).
The point is that these actions do not satisfy (i), namely they contain none of non-nearest-site hoppings necessary for Lorentz covariant excitations of fermions as discussed in Sec.~\ref{sec:features}.
As a consequence, although they preserve the discrete symmetry for a continuum limit, they inevitably produce unphysical fermions as seen above.
These results are consistent with the argument discussed in Sec.~\ref{sec:features} that the requisite discrete symmetry for a good continuum limit is incompatible with physical minimal-doubling actions on hyperdiamond lattices.

\subsection{Other fermions}

We then consider the lattice action based on the hyperdiamond lattice
with (i)non-nearest hoppings and (iii)lattice deformation while the spinor
structure is not twisted.
We call this Untwisted Creutz action.
The associated Dirac operator is given by $i \gamma_4 \to \gamma_4$ in Eq.~(\ref{Creutz_Dirac}) as
\begin{eqnarray}
  D(p)\  = 
 && (\sin p_1 + \sin p_2 - \sin p_3 - \sin p_4) i \gamma_1 \nonumber \\
 &+& (\sin p_1 - \sin p_2 - \sin p_3 + \sin p_4) i \gamma_2 \nonumber \\
 &+& (\sin p_1 - \sin p_2 + \sin p_3 - \sin p_4) i \gamma_3 \nonumber \\
 &+& B (4C - \cos p_1 - \cos p_2 - \cos p_3 - \cos p_4) \gamma_4. 
 \label{utCreutz_Dirac}
\end{eqnarray}
Thus we obtain minimal-doubling fermions for $1/2<C<1$ as Creutz action.
However, since $\gamma_4$ is not anti-hermitian, covariance of them is broken as
$i\vec{\gamma}\cdot\vec{k}+\gamma_4 k_4$.
As the case of Creutz action discussed in section \ref{sec:Creutz},
Untwisted Creutz action with the exact hyperdiamond spinor vectors is
given by the condition $C=1$. 
Then the same unphysical fermion $i \vec{\gamma} \cdot \vec{k}$ is
obtained at $p=0$ as Creutz fermion with $C=1$.

The remaining lattice fermions are BBTW fermion and Dropped Creutz fermion with the lattice deformation.
Since they include only nearest neighbor hoppings, both of
$i\gamma$-terms and $\gamma\gamma_5$-terms arise in the action, and thus
Nielsen-Ninomiya's no-go theorem cannot be applied to them as discussed before.
All these fermions implies the necessity of the three conditions for physical fermions on the hyperdiamond lattice.

\subsection{Appended Creutz fermion}\label{sec:appended}

In this section we propose a new lattice action, which also possesses physical minimal-doubling fermions as the case of the original Creutz action.
It is pointed out in section \ref{sec:features} that Creutz action
includes non-nearest-site hopping terms but nearest unit cell hoppings.
However, as shown in Fig.~\ref{honeycomb_nonnearest}, all of nearest unit cell hoppings are not included by Creutz action, and we now consider a new lattice action including all of them,
\begin{eqnarray}
  S_{\mathrm{aC}} & = & S_{\mathrm{C}} + \frac{1}{2} \sum_x \sum_{\mu<\nu} \Big[ \bar{\phi}_{x-\bm{a}_\mu+\bm{a}_\nu} \Sigma \cdot \left( \bm{s}^\mu - \bm{s}^\nu \right) \chi_x - \bar{\chi}_{x+\bm{a}_\mu-\bm{a}_\nu} \Sigma \cdot \left( \bm{s}^\mu - \bm{s}^\nu \right) \phi_x \nonumber \\
 && \hspace{6.5em} - \bar{\phi}_{x+\bm{a}_\mu-\bm{a}_\nu} \bar{\Sigma} \cdot \left( \bm{s}^\mu - \bm{s}^\nu \right) \chi_x + \bar{\chi}_{x-\bm{a}_\mu+\bm{a}_\nu} \bar{\Sigma} \cdot \left( \bm{s}^\mu - \bm{s}^\nu \right) \phi_x \Big]. \label{appended_action}
\end{eqnarray}
We will call this new action Appended Creutz action.
Here the hopping vectors $\bm{a}_{\mu}$ are obtained from the reciprocal vectors in the momentum space, which differ slightly from those of the original Creutz action.

The additive contribution to the Dirac operator is given by
\begin{eqnarray}
D'(p)\ = && 2 ( \sin p_{13} + \sin p_{14} + \sin p_{23} + \sin p_{24} ) i\gamma_1 \nonumber \\
 & + & 2 ( \sin p_{12} + \sin p_{13} - \sin p_{24} - \sin p_{34} ) i\gamma_2 \nonumber \\ 
 & + & 2 ( \sin p_{12} + \sin p_{14} - \sin p_{23} + \sin p_{34} )
  i\gamma_3
 \label{appended_Dirac}
\end{eqnarray}
where we define $p_{12} = p_1 - p_2$, etc.
The coefficient of $i\gamma_4$ is constantly zero and the total Dirac operator is obtained as the sum of
(\ref{Creutz_Dirac}) and (\ref{appended_Dirac}).
Thus it gives the same minimal-doubling poles at $p = \pm (\tilde p, \tilde p, \tilde p, \tilde p)$ with $\cos \tilde p = C$ and the minimal-doubling condition $1/2 < C < 1$ as the case of Creutz action.
Because the Dirac operator (\ref{appended_Dirac}) is written by only $i\gamma$-terms, the physical minimal-doubling fermions are obtained from (\ref{appended_action}).

This action satisfies all of three conditions discussed in section \ref{sec:features}, the non-nearest-site hoppings, the spinor twist and the lattice deformation.
As a result, it lacks the sufficient discrete symmetry to suppress the redundant operators generated by the loop corrections\cite{Capitani:2009yn}.
Actually although this Dirac operator (\ref{appended_Dirac}) includes
only $\gamma_1$, $\gamma_2$ and $\gamma_3$ terms, fermionic excitations
from the pole of this operator are the same as those of the original one
(\ref{Creutz_Dirac02}) up to a factor.
After all, Appended Creutz action is quite similar to the original Creutz action, and thus it suggests stability of the minimal-doubling poles to some kinds of perturbations.

\begin{table}[tp]
\begin{center}
\begin{tabular}{ccccc}
 \scalebox{.7}{Actions} & \scalebox{.7}{(i) Non-nearest hopping} & \scalebox{.7}{(ii) Spinor twist} & \scalebox{.7}{(iii) Lattice deformation} & \scalebox{.7}{Poles} \\ \hline
 \scalebox{0.7}{Creutz} & $\bigcirc$ & $\bigcirc$ & $\bigcirc$ & \scalebox{.7}{physical
		 minimal-doubling $p\not=0$} \\
 \scalebox{0.7}{Appended Creutz} & $\bigcirc$ & $\bigcirc$ & $\bigcirc$ & \scalebox{.7}{physical
		 minimal-doubling $p\not=0$} \\
  & $\bigcirc$ & $\bigcirc$ & $\times$ & \scalebox{.7}{unphysical $p=0$} \\
 \scalebox{0.7}{Untwisted Creutz} & $\bigcirc$ & $\times$ & $\bigcirc$ &
		 \scalebox{.7}{unphysical  minimal-doubling $p\not=0$} \\
  & $\bigcirc$ & $\times$ & $\times$ & \scalebox{.7}{unphysical $p=0$} \\ 
  & $\times$ & $\bigcirc$ & $\bigcirc$ & \scalebox{.7}{unphysicals $p\not=0$} \\
 \scalebox{0.7}{Dropped Creutz} & $\times$ & $\bigcirc$ & $\times$ &
		 \scalebox{.7}{unphysical $p=0$} \\
  & $\times$ & $\times$ & $\bigcirc$ & \scalebox{.7}{unphysicals $p\not=0$} \\
 \scalebox{0.7}{BBTW} & $\times$ & $\times$ & $\times$ & \scalebox{.7}{unphysicals
		 $p\not=0$ and physical $p=0$} 
\end{tabular}

\end{center}
\caption{List of lattice fermions based on the hyperdiamond lattice. The
 lattice actions including only physical modes are Creutz and Appended
 Creutz action, which are the minimal-doubling action. BBTW action has a
 physical and some unphysical poles. The others do not possess any
 physical poles. The fermions with the sufficient discrete symmetry are
 BBTW fermion and Dropped Creutz fermion.}
\label{3features}
\end{table}

At last, we have investigated all of lattice fermions in the context of
the three features as listed in Table \ref{3features}.
Furthermore, we have proposed a new lattice action, called Appended
Creutz action, which gives the physical minimal-doubling fermions.
After all, the lattice actions giving only physical fermions are the
original Creutz action and Appended Creutz action.
Other actions have some problems such as including both of $i\gamma$ and
$\gamma\gamma_5$ or breaking covariance.

While we have discussed free lattice fermions, we now remark effects of
gauge interaction.
As discussed in preceding studies, since one of the spacetime direction
is specified in lattice actions we have considered, we have to
renormalize the light of speed to resolve the anisotropy generated by
interactions.
Thus it is expected that the Lorentz
covariance is just modified, but not broken through gauge interactions
and quantum corrections because physical fermions discussed here actually
satisfy all of the conditions for Nielsen-Ninomiya's theorem.
On the other hand, it is slightly meaningless to consider quantum
corrections for the lattice actions including unphysical fermionic modes.

We then comment on chiral charge of the lattice fermions.
As the case of Creutz fermion, the physical minimal-doubling fermions we
have discussed possess the sublattice structure of the hyperdiamond
lattice, and thus their total chiral charge becomes zero.
However the index of the Dirac operator, which is interpreted as that of the
real vector field in the context of Poincar\'e-Hopf theorem, is
ill-defined if both of $i\gamma$ and $\gamma\gamma_5$ are included.

\section{A novel construction of minimal-doubling fermion}\label{sec:rhombus}

As discussed in Sec.~\ref{sec:hyperdiamond_Creutz}, we can construct
Creutz fermion on the distorted hyperdiamond lattice.
However in this real-space construction, the interactions based on the vectors $\bm{a}_{\mu}$ (primitive vectors) stand for hoppings from one unit cell to another unit cell, not between hyperdiamond-lattice sites.
In addition there is no nontrivial hopping between two sites in the same unit cell except for on-site terms.
As seen from these reasons, the real-space construction of Creutz fermion on the hyperdiamond lattice is somewhat misleading although it gives intuitive understanding of the discrete symmetry of the action, and it is much more natural that the two sites in one unit cell is identified as a single site as seen in Fig.~\ref{honeycomb_rhombus}.
Based on this argument, in this section we explicitly show alternative
and more reasonable spatial construction of Creutz-type minimal-doubling
action on a deformed hypercubic {\it rhombus} lattice, which is, in a sense, a generalized version of Creutz action.

Comparing it with another well-known minimal-doubling fermion on hypercubic
lattice, called Wilczek fermion\cite{Wilczek:1987kw},
both of them include additive terms proportional to $\gamma_4$.
However the location of poles of Creutz's Dirac operator is
adjustable while that of Wilczek fermion is not.
In this section we also consider a modification of Wilczek action,
and construct an orthogonal lattice action with adjustable poles.

\subsection{Creutz-type lattice action}

We especially remark the translation symmetry of the real-spatial
lattice on which the action is defined, and show a novel
minimal-doubling fermion including Creutz fermion can be constructed on
a deformed hypercubic (rhombus) lattice only with nearest neighbor hoppings. 

\begin{figure}[tbp]
 \begin{center}
  \includegraphics[height=10em]{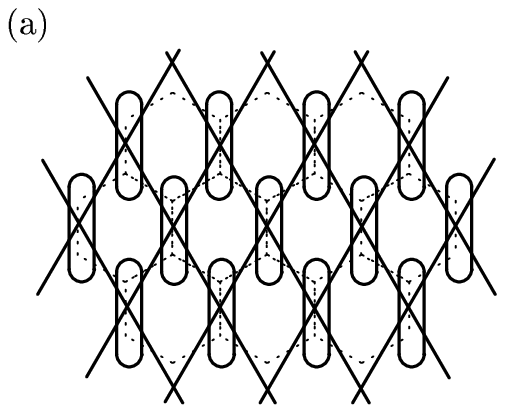} \quad
  \includegraphics[height=10em]{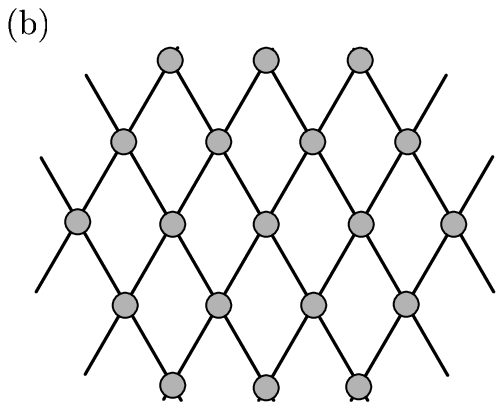}
 \end{center}
 \caption{(a) Translation symmetry of the honeycomb lattice. Unit cells
 consisting of $L$-node and $R$-node are encircled. (b)
 A deformed square (rhombus-like) lattice with the translation symmetry
 equivalent to that of the honeycomb lattice.}
 \label{honeycomb_rhombus}
\end{figure}

As shown in Fig.~\ref{honeycomb_rhombus}, we can consider rhombus-like
lattice which possesses the translation symmetry equivalent to that of the
hyperdiamond lattice.
Its spatial primitive vectors $\bm{a}_\mu$ will be found later, but we anyway introduce
a lattice action defined on the rhombus-like lattice, 
\begin{equation}
  S = \frac{1}{2} \sum_{x} \Bigg[\sum_{\mu=1}^4 \left(
						    \bar\psi_{x}\Gamma\cdot\bm{s}^\mu\psi_{x+\bm{a}_\mu}
						    -
						    \bar\psi_{x+\bm{a}_\mu}\bar\Gamma\cdot\bm{s}^\mu\psi_x
						   \right) +2i\,t\, \bar\psi_x \gamma_4 \psi_x \Bigg] \label{rhom_action01}
\end{equation}
with $\Gamma=(\vec\gamma, -i\gamma_4)$ and $\bar\Gamma=(\vec\gamma,
i\gamma_4)$.
Here $t$ stands for a free parameter to be fixed for minimal number of doublers.
This expression is similar to what is presented in
\citen{Buchoff:2008ei}, but in this paper we explicitly show alternative
spatial construction.
There are five spinor vectors in the previous lattice action
(\ref{Creutz_action}), but in this action only four vectors and an
on-site parameter instead of fifth vector.
This on-site term is a Wilson-like term with one specific
direction\cite{Karsten:1981gd,Wilczek:1987kw}.

If we apply (\ref{twisted_bond}) to spinor vectors $\{\bm{s}^\mu\}$,
this lattice action is reduced to Creutz action.
To show this, we investigate its momentum space structure of the action
(\ref{rhom_action01}).
Taking its Fourier transformation,
the associated Dirac operator is obtained as
\begin{equation}
 D(p) = i \sum_{\mu=1}^4 \zeta_\mu(p) \gamma_\mu \label{Creutz_Dirac03}
\end{equation}
where the coefficients of gamma matrices are
\begin{equation}
 \zeta_\mu(p) = \left\{ \begin{array}{ll}
		 \sum_{\nu=1}^4 \left(\bm{s}^\nu\right)_\mu \sin p_\nu  &
		 ( \mu=1,2,3 )\\
		t - \sum_{\nu=1}^4 \left(\bm{s}^\nu\right)_\mu \cos
		 p_\nu & (\mu = 4)
			\end{array} \right. ,
\end{equation}
and $\left(\bm{s}^\nu\right)_\mu$ represents $\mu$-th components of
$\bm{s}^\nu$.
In order that this Dirac operator has minimal-doubling poles at $p =
\pm(\tilde p, \tilde p, \tilde p, \tilde p)$ with $\tilde p>0$, spinor vectors should
satisfy 
\begin{equation}
 \sum_{\nu=1}^4 \left(\bm{s}^\nu\right)_\mu = \left\{\begin{array}{cl}
		 0 & (\mu=1,2,3) \\
		t / \tilde C & (\mu=4)				      
						     \end{array}\right. , \quad
 \tilde C = \cos \tilde p.
\end{equation}
It is easy to show that (\ref{twisted_bond}) actually satisfy this
condition, and to relate this Dirac operator (\ref{Creutz_Dirac03}) to
Creutz's Dirac operator (\ref{Creutz_Dirac}) we choose $\tilde C = C$
and $t=4BC$. 

To study fermionic excitations around poles, we expand the coefficients
of gamma matrices around poles with $p_\mu=\tilde p + q_\mu$,
\begin{equation}
 \zeta_\mu(p) = \left\{ \begin{array}{ll}
		 \tilde C \sum_{\nu=1}^4 \left(\bm{s}^\nu\right)_\mu
		  q_\nu + \mathcal{O}(q^2) &
		  (\mu=1,2,3) \\
		\tilde S \sum_{\nu=1}^4 \left(\bm{s}^\nu\right)_\mu q_\nu + \mathcal{O}(q^2) & (\mu = 4)
			\end{array} \right.
\end{equation}
with $\tilde S=\sin \tilde p$.
Thus reciprocal vectors are given by
\begin{equation}
 \left(\bm{b}^\nu\right)_\mu = \left\{ \begin{array}{ll}
		 \tilde C \left(\bm{s}^\nu\right)_\mu &
		  (\mu=1,2,3) \\
		\tilde S \left(\bm{s}^\nu\right)_\mu & (\mu = 4)
			\end{array} \right. .
\end{equation}
At this stage, we obtain primitive (hopping) vectors $\{\bm{a}_\mu\}$ of the
lattice by the condition $\bm{a}_\mu\cdot\bm{b}^\nu=\delta_\mu^{\nu}$.
In general, the lattice becomes non-orthogonal, deformed hypercubic lattice as
shown in Fig.~\ref{honeycomb_rhombus}.

An advantage of this action is of course that we can easily introduce
gauge fields by link variables on bonds of the rhombus-like lattice.
The fermionic part of the lattice action with gauge fields and a mass-term
is given by
\begin{equation}
  S = \frac{1}{2} \sum_{x,\mu} \left[
						    \bar\psi_{x}\Gamma\cdot\bm{s}^\mu
						    U_{x,\mu}\psi_{x+\bm{a}_\mu}
						    -
						    \bar\psi_{x+\bm{a}_\mu}\bar\Gamma\cdot\bm{s}^\mu
						    U^\dag_{x,\mu}\psi_x
						   \right] + \sum_x \left[ M\bar\psi_x\psi_x  +it \bar\psi_x \gamma_4
		     \psi_x \right].
\end{equation}
Note that link variables are also represented in non-orthogonal
coordinates.
Then we can introduce the action of gauge fields as the plaquette action on
the deformed hypercubic lattice.

\subsection{Orthogonal lattice action}

We have discussed much simpler and generalized expression of Creutz-type
minimal-doubling lattice fermion.
Then we claim minimal-doubling poles are due to the modification of the
lattice action by introducing on-site term which is proportional to
$\gamma_4$.
In this sense, its minimal-doubling mechanism is similar to Wilczek action.
However in the case of Creutz action, the location of poles is adjustable while
that of Wilczek action is not.
We then construct an orthogonal lattice action whose poles are
adjustable.
It will be the simplest form of Creutz-type lattice action, and
useful for us to understand its structure.

We now construct a lattice action on an orthogonal lattice,
\begin{equation}
  S = \frac{1}{2} \sum_{x,\mu}
 \left[ \bar\psi_x \gamma_\mu \psi_{x+\bm{a}_\mu}
  - \bar\psi_{x+\bm{a}_\mu}\gamma_\mu \psi_x
 \right]
 + \frac{i}{2}r \sum_x \left[
 2(3+t)\bar\psi_x\gamma_4\psi_x - \sum_{\mu=1}^4
 \left(
  \bar\psi_x \gamma_4 \psi_{x+\bm{a}_\mu}
  + \bar\psi_{x+\bm{a}_\mu}\gamma_4 \psi_x
 \right) \right],
\end{equation}
where $t$ and $r$ are free positive parameters to be fixed for minimal number of doublers.
The explicit expression of primitive vectors will be shown later.
The associated Dirac operator is given by
\begin{equation}
 D(p) = i \sum_{\mu=1}^4 \gamma_\mu \sin p_\mu + i r \gamma_4
  \left[\sum_{j=1}^3\left(1-\cos p_j\right) + \left(t-\cos
		     p_4\right)\right]. \label{Dirac_orth}
\end{equation}
To obtain poles of this Dirac operator, we take coefficients of
$\gamma$'s to be zero.
In the cases of $j=1,2,3$, we obtain
\begin{equation}
 \sin p_j = 0,\ \pi \quad (j=1,2,3).
\end{equation}
The coefficient of $\gamma^4$ reads
\begin{eqnarray}
  && \sin p_4 + r \left[\sum_{j=1}^3(1-\cos p_j) + (t-\cos p_4)\right]
  \nonumber \\
 & = & \sqrt{1+r^2} \sin(p_4-\alpha) + r (t+2N_\pi)
\end{eqnarray}
with $N_\pi=\#\{p_j=\pi,\ j=1,2,3\}$, $\cos \alpha = 1/\sqrt{1+r^2}$,
$\sin \alpha = r/\sqrt{1+r^2}$.
Thus the condition such that this Dirac operator induces only
minimal-doubling two poles is given by
\begin{equation}
 \left|\frac{r}{\sqrt{1+r^2}}\right|t < 1, \quad
 \left|\frac{r}{\sqrt{1+r^2}}\right|(t+2) > 1,
\end{equation}
and then minimal-doubling poles become $p = (0,0,0,p^{(\pm)}+\alpha)$
with $\sin p^{(\pm)} = -rt/\sqrt{1+r^2}$, $\cos p^{(\pm)} = \pm
\sqrt{(1+r^2(1-t^2))/(1+r^2)}$.

To obtain translation symmetry of the lattice action, we expand the
Dirac operator around a pole with momentum
$(q_1,q_2,q_3,q_4+p^{(+)}+\alpha)$ as
\begin{equation}
 D(p) = i\sum_{j=1}^3 \gamma_j q_j + i \sqrt{1+r^2(1-t^2)}\gamma_4 q_4 + \mathcal{O}(q^2).
\end{equation}
Therefore reciprocal vectors for this lattice action are represented by
\begin{equation}
 \left(\bm{b}^\nu\right)_\mu = \left\{ \begin{array}{cl}
		 \delta^\nu_\mu &
		  (\mu=1,2,3) \\
		 \delta^\nu_\mu \sqrt{1+r^2(1-t^2)} & (\mu = 4)
			\end{array} \right. ,
\end{equation}
and by the condition $\bm{a}_\mu\cdot\bm{b}^\nu=\delta_\mu^{\nu}$,
primitive vectors are obtained,
\begin{equation}
 \left(\bm{a}_\nu\right)_\mu = \left\{ \begin{array}{cl}
		 \delta_{\nu\mu} &
		  (\mu=1,2,3) \\
		\delta_{\nu\mu}/\sqrt{1+r^2(1-t^2)} & (\mu = 4)
			\end{array} \right. .
\end{equation}
It indicates the lattice action is constructed on orthogonal
lattice with one direction specified.
We now remark although both this Wilczek-type action and Bori\c{c}i action are
defined on an orthogonal lattice, they are not equivalent.

As the case of Creutz action, minimal-doubling poles induce two Dirac
fermions.
Although the Dirac operator (\ref{Dirac_orth}) has parity symmetry, the
action lacks sufficient discrete symmetry to remove redundant operators
generated by loop corrections.
Besides, when the fourth component of the
reciprocal vector is zero, covariance of the associated fermion is
broken and it becomes unphysical.

\section{Summary}\label{sec:summary}
In this paper we investigate minimal-doubling fermion actions on deformed-hyperdiamond lattices, with emphasis on the real-space construction of them and the correct excitations from poles of propagators, then generalize them to an action on a rhombus lattice.

In Sec.~\ref{sec:hyperdiamond_Creutz} we propose the spatial construction of Creutz fermion action on a deformed hyperdiamond lattice, where the hopping vectors (or the primitive vectors) are consistently determined by the reciprocal vectors in the momentum space.
It means that the spatial lattice structure, on which fermions live, depends not only on the spinor vector $\bm{s}^{\mu}$ but also on the form of the action itself.
Based on this construction we give a condition for the action to be
defined on the exact hyperdiamond lattice in the real space while a condition for the poles of propagators to be located on the hyperdiamond-lattice sites is proposed in \citen{Creutz:2007af}.

In Sec.~\ref{sec:features} we investigate the conditions for a hyperdiamond-lattice action to produce physical or Lorentz-covariant excitations from poles of fermion propagators, which actually Creutz fermion satisfies. 
Then it is pointed out that the non-nearest-site (but nearest-unit-cell) hoppings are essential for the correct excitations from the poles of doublers:
If the action on the hyperdiamond lattice does not contain any non-nearest-site hopping,
it shall yield fermions with unphysical excitations since the associated Dirac operator inevitably includes both of  $i\gamma$- and $\gamma \gamma_{5}$-terms while Nielsen-Ninomiya's no-go theorem assumes either of vector or axial-vector functions.
This fact implies that the requisite discrete symmetry of the action for a good continuum limit is incompatible with ``minimal-doubling'' with correct excitations on hyperdiamond lattices as firstly discussed in \citen{Bedaque:2008jm}.  
All the fermion actions we have discussed in Sec.~\ref{sec:examples} back up this incompatibility.
In the section we also study a related minimal-doubling fermion called ``Appended Creutz fermion'', which contains all the nearest neighbor interactions in terms of unit cells.

As discussed in Sec.~\ref{sec:rhombus}, we can construct Creutz-type minimal-doubling actions more naturally on a deformed hypercubic lattice, instead of a hyperdiamond lattice. 
We propose a class of minimal-doubling fermion actions defined on a rhombus lattice.
In a sense, this class of the actions is a generalization of Creutz-type actions since it reduces to the original Creutz action by choosing the parameters appropriately.
Based on this alternative and reasonable spatial construction of minimal-doubling actions,
the link variables are easily introduced.
In the section we also introduce a two-parameter class of Wilczek-type
minimal-doubling actions,
which is the most simple form of Creutz-type action.

As a future work we will search for a general and unified form of minimal-doubling actions, which can reduce to all the known minimal-doubling actions including Wilczek fermion and Creutz fermion.
This kind of the actions, if exists, will reveal more on the incompatibility between ``minimal-doubling'' and the requisite discrete symmetry for a good continuum limit.

\section*{Acknowledgments}
We would like to thank T. Onogi for reading the manuscript and useful discussions.
We also thank S. Aoki, M. Creutz and Y. Kikukawa for valuable comments.
TM is supported by Grant-in-Aid for the Japan Society for Promotion of Science (JSPS) Research Fellows.


\providecommand{\href}[2]{#2}\begingroup\raggedright\endgroup

\end{document}